\documentclass[prl,aps,multicol]{revtex4}
\usepackage{epsfig}
\newcommand{\sym}{ {\cal S}}
\newcommand{\zst}{z^*}
\newcommand{\Zst}{Z^*}
\newcommand{\norm}{{\cal N}}
\newcommand{\beq}{\begin{equation}}
\newcommand{\eeq}{\end{equation}}
\newcommand{\barray}{\begin{eqnarray}}
\newcommand{\earray}{\end{eqnarray}}

\begin{document}
\title{Spin-s wavefunctions with algebraic order}
\author{Onuttom Narayan and B. Sriram Shastry }
\affiliation{
Department of Physics, University of California, Santa Cruz, CA 95064}

\date{\today}

\begin{abstract}
We generalize the Gutzwiller wavefunction for $s={1\over2}$ spin chains
to construct a family of wavefunctions for all $s>{1\over 2}$. Through
numerical simulations, we demonstrate that the spin spin correlation
functions for {\it all\/} $s$ decay as a power law with logarithmic
corrections. This is done by mapping the model to a classical statistical
mechanical model, which has coupled Ising spin chains with long range
interactions. The power law exponents are those of the Wess Zumino Witten
models with $k=2s.$ Thus these simple wavefunctions reproduce the spin
correlations of the family of Hamiltonians obtained by the Algebraic
Bethe Ansatz.
\end{abstract}

\maketitle

\section{Introduction}
Quantum spin chains with power law correlations are of great current
interest. Starting with the Bethe chain~\cite{Bethe}, it has been
known for a long time that half integer spin systems generically have
spin correlation functions that decay as the inverse of the distance
between the spins. For integer spin chains, the correlation functions
are generically expected to decay exponentially, due to the Haldane
conjecture~\cite{haldane}. However, it is also known that special models
can be constructed for which the spin correlation functions decay as a
power law for all (including integer) spin $s$, with an exponent that
depends on $s.$ These models are realized from the Bethe Ansatz in its
algebraic form. The work of the Leningrad school has provided concrete
realizations, e.g. the model of Takhtajan and Babudjan~\cite{babudjan}
for spin 1. It has also provided a general technique for obtaining the
Hamiltonian for all values of spin, starting from the Algebraic Bethe
Ansatz~\cite{kulish}.

The coefficients of the biquadratic and other terms need to be specially
chosen to obtain these models. Alternatively, the Algebraic Bethe
Ansatz method automatically generates these Hamiltonians.  The general
construction of a field theory for  these models was undertaken by
Affleck~\cite{affleck}, who showed that these models belong to the Wess
Zumino Witten(WZW) class. These models are characterized by a single
parameter $k,$ which determines the exponents. In this case, $k$ turns out
to be $2 s$ for spin $s$. From the known behaviour of the WZW theories,
Affleck showed that the spin spin correlations are asymptotically of
the form
\beq 
|\langle\vec{S}_i . \vec{S}_j\rangle| \sim |i - j|^{-3/(2 s + 2)}.  
\label{affl}
\eeq
For any $s,$ the field theory has a marginal operator in the renormalization
group sense. This gives rise to a multiplicative factor~\cite{marginal} of 
$[A + B\ln|i - j|]^{1/2}$ in the correlation function of Eq.(\ref{affl}), 
unless the bare value of the marginal operator is fortuitously equal to its 
fixed point value~\cite{haldane2,shastry}.

An alternative recent line of thought has been to study explicit wave
functions. In particular, the Gutzwiller wave function in one dimension
has been very successful. The Gutzwiller wave function takes the free
Fermion determinantal wavefunction and retains all configurations
with single occupancy of electrons. Thus at half filling, i.e. one
electron per site, it yields a wave function that has only spin
degrees of freedom. This insulating spin wavefunction inherits the
power law correlations of the parent free Fermi wavefunction, albeit
with some renormalization of the values of the exponents. The work of
Haldane~\cite{haldane2} and Shastry~\cite{shastry} and others~\cite{others}
showed that the wavefunction is the exact ground state of a long ranged
$s=1/2$ Heisenberg model that is in the same universality class as the
Bethe chain, i.e. has the same correlation exponents.

In this paper, we propose a natural extension of the Gutzwiller
wavefunction to obtain wavefunctions for {\it all\/} $s,$ which (based
on our numerical results) seem to have power law correlations of the
form in Eq.(\ref{affl}). Thus we have constructed for wave functions
the analog of the Algebraic Bethe Ansatz method for Hamiltonians: a
prescription that yields the non-generic correlations of Eq.(\ref{affl})
automatically, i.e. without any fine tuning of parameters. This paper
builds on ideas presented earlier in Ref.~\cite{shastry2}.

We start with $2s$ copies of the Gutzwiller wave function, and use the
principle of symmetrization to produce an angular momentum $J= s$
wave function, i.e. one where each lattice site has a spin $s$ degree
of freedom. Symmetrization is a well known procedure in angular momentum
theory, where one generates the states of a spin $J$ system by taking $2
J$ copies of spin half states and projecting out all states that are not
fully symmetric in the $2J$ spin constituents. This procedure clearly
generates wave functions for particles of angular momentum $s.$ Based
on the experience with $s=1/2,$ one would hope that these wavefunctions
might also inherit the power law correlations of the parent free Fermi
gas, again with some as yet undetermined renormalization of values
of exponents.

We are able to perform the symmetrization of  this wave function
explicitly, using the elegant formalism of spin coherent states. We
further study the properties of this wave function using numerical
techniques. When the wave function is squared, it can be conveniently
interpreted as a statistical mechanical model of $4s$ parallel chains
of logarithmically interacting particles with certain couplings, i.e. a
generalization of the Wigner Dyson Coulomb problem.  Our numerical results
for $s=1, 3/2, 2 $ are presented here. Together with the analytically
known results for $s=1/2,$ they support the remarkable conclusion that
these wave functions provide a {\em lattice realization} of the WZW
models with $k= 2s$.

We are unable to address the issue of finding a Hamiltonian for which
the states here are exact ground states, but have wave functions that
are explicit and rather beautiful. This seems to be just the opposit
situation of the Algebraic Bethe Ansatz models~\cite{kulish} for higher
spin, where the Hamiltonians are relatively straightforward, but the
wave functions are highly complicated.

Earlier numerical work by Shastry~\cite{shastry2} on the same wave
function for $s=1,$ with relatively shorter spin chains, gave results
that were consistent with the same exponents as found here, but lacked
the resolution to determine the logarithmic corrections to the leading
behavior. We note that spin coherent states were used for a similar
mapping of spin wave functions to statistical mechanical problems
in the work of Arovas, Auerbach and Haldane~\cite{arovas} as well as
Affleck, Kennedy, Lieb and Tasaki (AKLT), who studied the $s=1$ model of
AKLT~\cite{aklt}, for which correlations are exponential. The calculation
of the correlations is considerably easier than in the cases studied
here. AKLT's method of solution is also a very nice application of the
idea of symmetrization that we use in the present paper. Analytical 
and numerical results for the spin-half chain, with the corresponding
statistical mechanical model studied as a function of temperature, have
been presented earlier~\cite{jphysa}.

\section{Spin Coherent States, Symmetrization and the Gutzwiller wave function}
\subsection{Spin Coherent States}
We begin by recapitulating the salient definitions of spin coherent
states. There are several related variants of these, and each has slightly
different advantages. Let us consider a single site. The functions
introduced by Radcliffe~\cite{radcliffe} are defined for angular momentum
$J$ as 
\barray
|\Omega\rangle &=& e^{[S^{-} 
\exp( i \phi) - S^{+} \exp( - i \phi) ]\theta/2 }|J\rangle \nonumber \\
&=& \cos^{2J}(\theta/2) 
e^{\tan(\theta/2) \exp(i \phi) S^{-} } |J\rangle \nonumber \\
&=&\sum_{n=0}^{2 J} \{^{2J}C_{n}\}^\frac{1}{2} \cos^{2J-n}(\theta/2)
\sin^n(\theta/2) e^{i n \phi} |J-n\rangle. 
\earray
Radcliffe also introduced a related set of states,
\beq
|z\rangle_R= e^{ z S^- - z^* S^+} |J\rangle.
\eeq
These are normalized and related to the first set by setting $z
\rightarrow \tan(\theta/2) e^{i \phi}.$ 
Another set of coherent states were found to be very convenient~\cite{sar}
for the purpose of obtaining differential operator representations of
spin operators, e.g. $S^z \rightarrow (s - z^* \partial/\partial z^*)$
in the space of ``wave functions'' $\langle z| \psi\rangle.$ These
coherent states were the unnormalized  states:
\beq
|z\rangle= e^{ z S^{-}} |J\rangle = \sum_{n=0}^{2J}  \
\{^{2J}C_{n}\}^\frac{1}{2} (z)^n |J - n \rangle. 
\label{unnormalized}
\eeq  
with $|z\rangle_R= (1+|z|^2)^{-J} \;  |z\rangle$. 
The relationship between the wave functions 
$\psi(z^*) \equiv \langle z|\psi\rangle$, $\psi_R(z^*) \equiv _R\langle z|\psi\rangle$ 
and $\psi(\Omega^*) $ follows from the relationship between the basis
functions, we note the relation needed later:
\beq
\psi(\Omega^*)= \cos^{2J}(\theta/2) \; \; 
\psi(z^*) \; \mbox{with~ }\; \; 
z^* \rightarrow \tan(\theta/2) e^{- i \phi}. 
\label{convert}
\eeq
The functions $\psi(z^*)$ are polynomials in $z^*$ of maximal degree $
2J$, and lend themselves to very simple ``symmetrization'' rules that
we discuss in the next subsection of this paper.

Next, we need to construct the rules for obtaining averages of variables
in states.
We begin by noting the
diagonal representation of an arbitrary operator $A$ \cite{arecchi}
followed by its average as:
\beq
A =  \frac{(2 J+1)}{ 4 \pi} \int \;d\Omega\; 
a(\Omega) |\Omega\rangle\langle\Omega|  
\label{diagonalrep} \\
\eeq
where the prefactor has been chosen so that $a(\Omega) = 1$ for the
identity operator. To compute averages, we write
\barray
\langle\psi|A|\psi\rangle &=&  \frac{(2 J+1)}{ 4 \pi} \int \; d\Omega \; 
a(\Omega)  \langle\Omega|\psi\rangle\langle \psi|\Omega\rangle \nonumber \\
&=&  \frac{(2 J+1)}{ 4 \pi} \int  \; d\Omega \;  
a(\Omega) |\psi(\Omega)|^2.
\label{avgs}
\earray
Thus, given a wave function $\psi(\Omega^*)\equiv
\langle\Omega|\psi\rangle,$ we can find the expectation value of any
operator if its $a(\Omega)$ is known. For future reference, we note that
$a(\Omega)$ for the operator $S_z$ is $(J+1)\cos\theta.$
An important corollary that we will use is that if spin coherent states
are constructed for every site in a lattice, the expectation value of an
operator $\langle A_i B_j\rangle$ can be found by using the corresponding
weight function $a(\Omega_i) b(\Omega_j)$ in the integrals, where
$a(\Omega)$ and $b(\Omega)$ are the functions for isolated sites. This
will be used for the spin correlation function.

\subsection{Symmetrization}
 Let us consider a simple case of two spin $\frac{1}{2}$ particles with 
\beq
|\phi\rangle = \alpha_1 \beta_2.
\eeq
Symmetrization is best understood from its action on states, so in the
present case:
\beq
 \sym |\phi\rangle= \frac{1}{2} (  \alpha_1 \beta_2 +  \alpha_2 \beta_1).
\eeq
On more general functions of many copies of spin half, its action is
similarly defined, namely find the fully symmetric combination generated
from  a seed state and divide by the total number of generated states.
We next deduce  the rule for symmetrizing $2 s$ copies of spin 1/2 in
the space of coherent state wavefunctions.  It turns out that  the
most effecient way is to work with the unnormalized coherent states
Eq(\ref{unnormalized}) where the wave functions are just polynomials
in $\zst$.  Let us label the $2 s$ copies of spin 1/2 by $\zst(\alpha)$
and the resulting spin s variables by $Z$, so that a coherent states of
the direct product states and the final spin s state are:
\barray
|\{z(1),z(2)\ldots z(2s)\}\rangle& = & \prod_{\alpha=1}^{2s}  
( |1/2\rangle_\alpha + z(\alpha) |-1/2\rangle_\alpha) \\
|Z\rangle&=& \sum_{m=0}^{2 s}   \{^{2s }C_{m}\}^\frac{1}{2} (Z)^m |2 s - m\rangle.
\earray
Generic states in the direct product space and the final spin s space
are represented by $|\phi\rangle$ and $|\Phi\rangle$ respectively, and the role of
symmetrization is to map the former into the latter as
\barray
\sym |\phi\rangle & =&| \hat{\phi}\rangle \nonumber \\
 |\hat{\phi}\rangle & \leftrightarrow & |\Phi\rangle.
\earray
The question we address is: given a state $|\phi\rangle$, what is the resulting
state $ |\Phi\rangle$ for the spin $2s$ particle. The answer turns out to be
remarkably simple in the unnormalized coherent state basis. If we denote
\barray
f(\{\zst(1),\zst(2)\ldots\zst(2s)\}) & = & 
\langle \{\zst(1),\zst(2)\ldots\zst(2s)\} |\phi\rangle \nonumber \\
\hat{f}(\{\zst(1),\zst(2)\ldots\zst(2s)\}) & = & 
\langle \{\zst(1),\zst(2)\ldots\zst(2s)\}  \sym |\phi\rangle \nonumber \\
F(Z^*)&=& \langle Z^*|\Phi\rangle,
\earray
we find ( and demonstrate below) that 
\beq
F(Z^*) = f(Z^*,Z^*\ldots Z^*).
\eeq 
This implies  that by    ignoring  the distinction between the different
copies of the spin 1/2 and replacing every occurrence of $z^*(\alpha)$ by
$Z^*$  gives us a coherent state representative of the spin s particle.
This result is obvious for the special ( symmetric) cases of $f=1$ and
$f =\prod \zst(\alpha)$, but not so obvious  for other cases,   since
one has to rule out  possible nontrivial dependence on the degree of
the polynomial.

We now give a brief proof of this assertion. It suffices to consider
the general case of a polynomial of degree $r$, thus
\barray
f &=&  \zst(1) \zst(2)\ldots \zst(r) \nonumber \\
\hat{f}&= & \frac{ r! (2s -r)!}{ 2s!} 
\sum_{ 1 \leq i_1 \langle i_2 \ldots\langle i_r \leq (2s)} \zst(i_1) \zst(i_2)\ldots\zst(i_r). 
\earray 
The state $\hat{f}$ clearly is proportional to the state deriving from
$(S^-)^r | J_z= 2s\rangle$, the proportionality constant is readily worked
out so
\beq
|\Phi\rangle= \frac{ ( 2s-r)!}{ (2s)!} (S^-)^r | 2s\rangle,
\eeq
hence
\barray
F(Z^*) & = & (Z^*)^r  \{ \frac{ ( 2s-r)!}{ (2s)!  r! } 
\langle 2s| (S^+)^r (S^-)^r | 2s\rangle \} \nonumber \\
&=&(Z^*)^r.
\earray
The last line follows on using the commutation relations of angular momentum.
 
\subsection{Gutzwiller type wave functions}
The Gutzwiller wave function for a one dimensional Fermi gas at half
filling is expressible in the form~\cite{shastry}
\barray
|\psi_G\rangle &=& \norm \prod_{j=1}^L 
[1 - n_{j,\uparrow} n_{j,\downarrow}] 
\prod_{|k|\leq k_F} c^\dagger_{k,\uparrow}c^\dagger_{k,\downarrow} |0\rangle 
\nonumber \\
&=& \norm \sum_{1\leq r_1 \langle\ldots\langle r_n\leq L} 
e^{\{i \pi \sum_j r_j \}} 
[ \prod_{k\langle j} \sin^2(\pi ( r_j- r_k)/L)  ] \; 
S_{r_1}^{-} S_{r_2}^{-}\ldots S_{r_n}^{-}\;\; | L/2\rangle
\earray
with $n=L/2.$ 
Here $c^\dagger_k$ is a creation operator with wavevector $k$ in
the original Fermionic representation, and the $[1 - n_{j,\uparrow}
n_{j,\downarrow}]$ factors ensure no double occupancy for any site $j.$
$S^-$ are the spin lowering operators in the spin representation that is
equivalent at half filling. $\norm$ is the normalization that we will
not specify till the end, since it cancels out in the evaluation of
the  correlations. For simplicity we have confined our considerations
to the case of $L/2$ overturned spins, so we are dealing with a global
singlet wave function, made up of L spin 1/2 particles.

We next consider 2s copies of this wave function, and project into the
spin $s$ sector at each site. In view of the discussion in the last
section, this is most easily done with the coherent state notation,
so the product  wave function is written down directly as
\barray
\Psi(\Zst_1, \Zst_2,\ldots\Zst_L)
& =&  \norm [  \sum_{1\leq r_1 \langle \ldots\langle r_\frac{L}{2}\leq L} 
e^{\{i \pi \sum_j r_j \}} [ \prod_{k<j} \sin^2(\pi ( r_j- r_k)/L)  ] \; 
\Zst_{r_1} \Zst_{r_2}\ldots\Zst_{r_{L/2}} ]^{2s}. \\
\earray
To obtain this result, we wrote $|\psi_G\rangle$ in terms of
$\zst(\alpha)$, multiplied $2s$ copies of this, and then symmetrized the
wavefunction by dropping the distinction between the different copies
or replicas. To reconstruct the wave function in the angular basis we
use Eq.(\ref{convert}) and write
\barray
\Psi(\Omega^*_1,\Omega^*_2\ldots\Omega^*_L)
&=& \prod_{j=1}^L\cos^{2s}(\theta_j/2)\Psi(\Zst_{r_1}, 
\Zst_{r_2}\ldots\Zst_{r_\frac{L}{2}}).
\earray
In order to make this more tractable, we introduce ``occupation numbers''
$\rho_j^\alpha$ which determine whether we get a $\cos(\theta_j/2)$
or $\sin(\theta_j/2)$ factor at a given site $j$ in a particular ``
replica'' $\alpha.$ Thus at each site we get a factor of
\barray
\mbox{factor} &=& \prod_{\alpha=1}^{2 s}[\cos(\theta_j/2) 
(1 - \rho_j^\alpha) +  
\sin(\theta_j/2) e^{- i \phi_j + i \pi j }  \rho_j^\alpha  ] \nonumber \\
&=&\cos^{2s}(\theta_j/2) \exp[{\sum_{\alpha=1}^{2s} \rho_j^\alpha 
\ln\{\tan(\theta_j/2)+ i \pi j - i \phi_j\}]}.
\earray
We thus write the wave function as
\beq
\Psi(\Omega^*_1,\Omega^*_2\ldots\Omega^*_L)= 
\norm \prod_{j=1}^L \cos^{2s}(\theta_j/2) \sum'_{\{ \rho_j^\alpha=0,1 \} }
\exp\left[\sum_{\alpha=1}^{2s}  
\left(\sum_{j<k} \rho_j^\alpha \rho_k^\alpha\ln\sin^2\frac{\pi (k - j)}{L} 
+\sum_j\rho_j^\alpha 
\ln\tan \frac{\theta_j}{2}+ i \pi j - i \phi_j  \right)\right] 
\label{psi}
\eeq
The sum over the occupancy integers $\rho^\alpha$ is constrained to
obey $\sum_n \rho_n^\alpha = L/2$ for each replica $\alpha$.
In the next section we continue the discussion of expectation values
and correlation functions, which require taking the modulus square of
this wave function.

\section{Coupled Ising representation}
The probability density is $ |\Psi(\{ \Omega^*_1\ldots\Omega^*_L\})|^2,$
so we need to multiply $\Psi$ in Eq.(\ref{psi}) with its complex
conjugate, leading to $4s$ replicas in all. The correlation function
$\langle s^z_j s_k^z\rangle$ can be found by calculating $(s+1)^2\langle
\cos\theta_j\cos\theta_k\rangle$ with this probability density, as noted
after Eq(\ref{avgs}). From the fact that $|\Psi\rangle$ is a singlet
state, this correlation function suffices to determine all components
$\langle s^\alpha s^\beta\rangle$. The angular variables $\theta$ and
$\phi$ can be integrated over, leaving only the $\rho_j$'s. The problem
reduces to one of interacting lattice gas particles.

Integration over the azimuthal angle $\phi_j$ can be done at each
site $j$, and gives a constraint that $\sum_\alpha \rho_j^\alpha =
\sum_{\alpha'} \rho_j^{\alpha'} \equiv  s- m_j$, where   $\alpha$
and $\alpha^\prime$ refer to replicas in $\Psi$ and $\Psi^*$
respectively. Note that the oscillating phase factor can be dropped in
view of the constraint from the azimuthal integration.

The integral over $\theta_j$ is next performed:
\begin{equation}
\int_0^\pi (\sin{\theta_j\over 2})^{2s - 2m} (\cos{\theta_j\over 2})^{2s + 2m}
\sin\theta_j d{\theta_j} = 
2 {{(s + m)! (s - m)!}\over{(2 s + 1)!}} = 2 W_s(m), 
\label{wsm}
\end{equation}
with the function $W_s(m)$ is defined by this equation. For the spin
autocorrelation function $\langle s^z_j s_k^z\rangle,$ we have to insert
an extra factor of $(s+1)\cos\theta$ in the $\theta$ integrals at $j$
and $k.$ It is easy to verify that
\begin{equation}
(s+1)\int_0^\pi (\sin{\theta_j\over 2})^{2s - 2m} 
(\cos{\theta_j\over 2})^{2s + 2m}
\cos\theta_j \sin\theta_j d{\theta_j} = 2 m W_s(m).
\label{magn}
\end{equation}

The final result is like a classical partition function:
\begin{equation}
Z = |\Psi(\{ \Omega^*_1\ldots\Omega^*_L\})|^2 
=  \norm \sum_{\rho_j^\alpha,\rho_k^{\alpha^\prime}}
\exp[\sum_{j<k} \rho_j^\alpha\rho_k^\alpha\ln\sin^2 (\pi (j - k)/L)
+ \sum_{j<k} \rho_j^{\alpha^\prime}\rho_k^{\alpha^\prime}
\ln\sin^2 (\pi (j - k)/L)]\prod_j W_s(m_j) 
\label{partitionfn}
\end{equation}
where $\alpha$ and $\alpha^\prime$ refer to replicas in $\Psi$ and
$\Psi^*$ respectively, $m_j  = s - \sum_\alpha \rho_j^\alpha$ and the
sum in the partition function $Z$ is subject to the constraints
\barray
\sum_j \rho_j^\alpha & =& \sum_j \rho_j^{\alpha^\prime} = L/2 \nonumber \\
\sum_\alpha\rho_j^\alpha & = & \sum_{\alpha^\prime} \rho_j^{\alpha^\prime}.
\label{constraint}
\earray
When calculating the spin autocorrelation function $\langle s_j^z
s_k^z\rangle,$ the summation in Eq.(\ref{partitionfn}) for $Z$ is
evaluated with an extra factor of $m_j m_k.$ 

It is convenient to change variables to $\sigma^\alpha_j = 2\rho^\alpha_j
- 1$ and $\sigma^{\alpha^\prime}_j = 1 - 2\rho^{\alpha^\prime}_j.$ We
then have $4 s$ coupled Ising chains ($2 s $ from $\Psi$ and $2 s $
from $\Psi^*$).  The constraint Eq.(\ref{constraint}) is equivalent to
the condition that the sum of the Ising spins (not to be confused with
the original quantum spins) across the different chains at any site must
be zero. In addition, we impose the condition that the sum of the spins
along any chain must be zero.  With this condition, the interactions
along any chain are
\begin{equation}
\sum_{j<k} {{1 + \sigma_j^\alpha}\over 2} {{1 + \sigma_k^\alpha}\over 2}
\ln\sin^2 (\pi (j - k)/L) = const + \sum_{j<k} {1\over 4}  \sigma_j^\alpha \sigma_k^\alpha
\ln\sin^2 (\pi (j - k)/L).
\end{equation}
The interactions along any chain are antiferromagnetic and logarithmic.
The interaction across the different chains at any site occurs, apart
from the constraint, through the $W_s(m)$ factor.

For the case of $s=1,$ there is an alternative form of the partition
function that is more convenient. With the constraint that the spins on
the four chains at any site must add up to zero, there are six possible
configurations at any site: $(1,1, -1, -1),$ $(1, -1, 1, -1),$ $(1,
-1, -1, 1),$ and the mirror images of these three. The magnetization
$m$ is 1 for the first and zero for the next two configurations. A
remarkable simplification occurs once we note that $\sum_\alpha
\sigma_j^\alpha\sigma_k^\alpha F(j - k)$ is equal to zero for any
function $F$ unless the configurations at the sites $j$ and $k$ are
either identical or mirror images. (In the sum over $\alpha$, the two
replicas from $\psi_s$ and the two from $\psi_s^*$ are included.) Thus
instead of four coupled Ising chains, the problem reduces to a six state
Potts model on one chain. The states are labelled by $q=\pm 1, 2, 3.$
The configuration $(1, 1, 1, 1)$ is labelled with $q=1,$ so that $m_j
= \pm 1$ when $q_j = \pm 1,$ and $m_j = 0$ when $q_j = \pm 2, \pm 3.$
The partition function is
\begin{equation}
Z \propto \sum_{q_1\ldots q_L}\exp[\sum_{j<k} (\delta_{q_j, q_k} -\delta_{q_j, - q_k})
\ln\sin^2 (\pi (j - k)/L)] \prod_j (1 +\delta_{q_j^2, 1}).
\label{Potts}
\end{equation}
The last factor comes from the fact that $W_s(\pm 1) = 2 W_s(0).$ Two
sites only interact with each other if their $q$'s are identical or
opposite. The condition that the total magnetization for each of the
four original chains must be zero reduces to the statement that
\begin{equation}
\sum_j \delta_{q_j, n} = \sum_j\delta_{q_j, -n}
\label{potts_constraint}
\end{equation}
for any $n.$

Potts models can similarly be constructed for $s > 1,$ although they
are more complicated: two sites $j$ and $k$ interact even when $q_j\neq
\pm q_k,$ and the condition from the total magnetization is weaker
than Eq.(\ref{potts_constraint}).  The numerical simulations reported
in the next section were conducted with both the coupled Ising and the
Potts representations.

\section{Numerical results}
Monte Carlo simulations were performed on the Ising and Potts
representations given by Eqs.(\ref{partitionfn}) and (\ref{Potts}). For
the Ising case, the constraint that the total spin along any chain
and for all replicas at a site must be zero prevents using single spin
flip dynamics. In a Monte Carlo move, two replicas $(\alpha,\beta)$ on
two adjacent sites $(j, k)$ were chosen at random. If the Ising spins
at these four locations satisfy $\sigma_j^\alpha = \sigma_k^\beta,$
$\sigma_j^\beta=\sigma_k^\alpha$ and $\sigma_j^\alpha\neq\sigma_j^\beta,$
it is possible to flip all the four spins simultaneously without violating
the constraint. The ratio of the probability of the flipped configuration
to the probability of the unflipped configuration is calculated, with
appropriate factors of $W_s(m)$ included if the move would change the
magnetization at the sites, i.e. if $\alpha \leq  2s < \beta$ or $\beta
\leq  2s < \alpha.$ The move is accepted or rejected using the standard
Metropolis criterion.

\begin{figure}
\centerline{\epsfxsize=4in \epsfbox{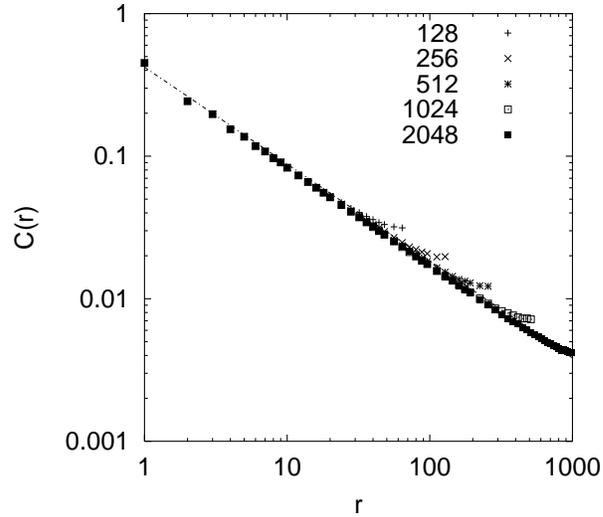}}
\caption{Log-log plot of $C(r),$ the magnitude of the spin autocorrelation
function, as a function of $r$ for $s=1.$ System sizes ranging from
$L=128$ to $L=2048$ are plotted. The error bars are comparable to the 
point size. The data is consistent with a power law
form, with exponent $-0.68.$ Since logarithmic corrections are expected,
a scaling collapse is not shown.
}
\label{fig1}
\end{figure}

For the Potts representation, two adjacent sites $(j,k)$ were chosen at
random, and $(q_j, q_k)$ were attempted to be changed. For the case of
$s = 1,$ Eq.(\ref{potts_constraint}) required that if $q_j + q_k\neq 0,$
the only possible move was to exchange them, while if $q_j + q_k = 0,$
one could attempt to replace them with either of the other two pairs
of $q$-values. For $s > 1,$ a table was constructed at the beginning
of the numerical simulation. For any pair of $q$-values $(q_1, q_2),$
the table listed all pairs $(q_1^\prime, q_2^\prime)$ that $(q_1, q_2)$
could change to, while respecting the magnetization constraints on the
underlying Ising spins. In any Monte Carlo step, this table was used to
randomly select an allowable move to attempt. For both the Ising and the
Potts representations, the long range logarithmic interaction down the
chains made calculating the probability of an attempted move an $O(L)$
long calculation for a chain of length $L.$ As a result, very large
values of $L$ could not be simulated.

\begin{figure}
\centerline{\epsfxsize=4in\epsfbox{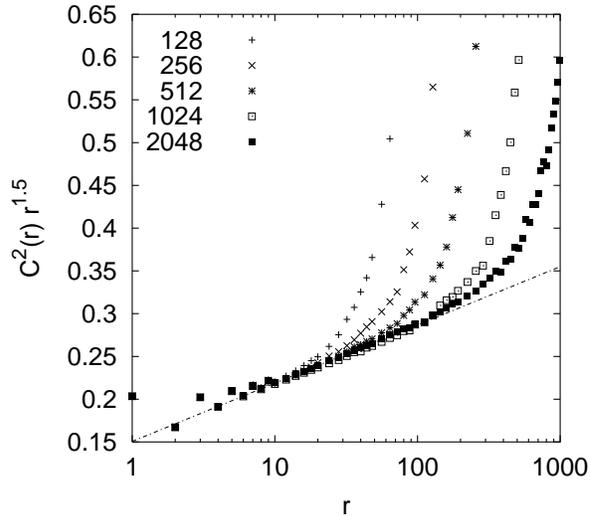}}
\caption{Plot of $C^2(r) r^{1.5}$ as a function of $\ln r$ for $s=1.$ This
is predicted to be a straight line, with finite size corrections. System
sizes from $L=128$ to $L=2048$ are plotted. The error bars are comparable 
to the point size.
}
\label{fig4}
\end{figure}

In both cases, error bars on the measured correlation function were
estimated by taking blocks of 30000 readings, calculating the average
correlation function within each block and then the inter-block
variance. Even if the individual readings are taken too frequently
and are therefore correlated, this procedure should be reliable so
long as the blocks are sufficiently large to be uncorrelated. It is
also useful to compare the variance of the block averages to the
variance of the individual readings within a block. The latter would
be 30000 times the former if the readings were uncorrelated. For
our simulations, the actual ratio ranged from about 30000 to 300,
confirming that in all cases the blocks are uncorrelated even when
the individual readings are not.

\begin{figure}
\centerline{\epsfxsize=4in\epsfbox{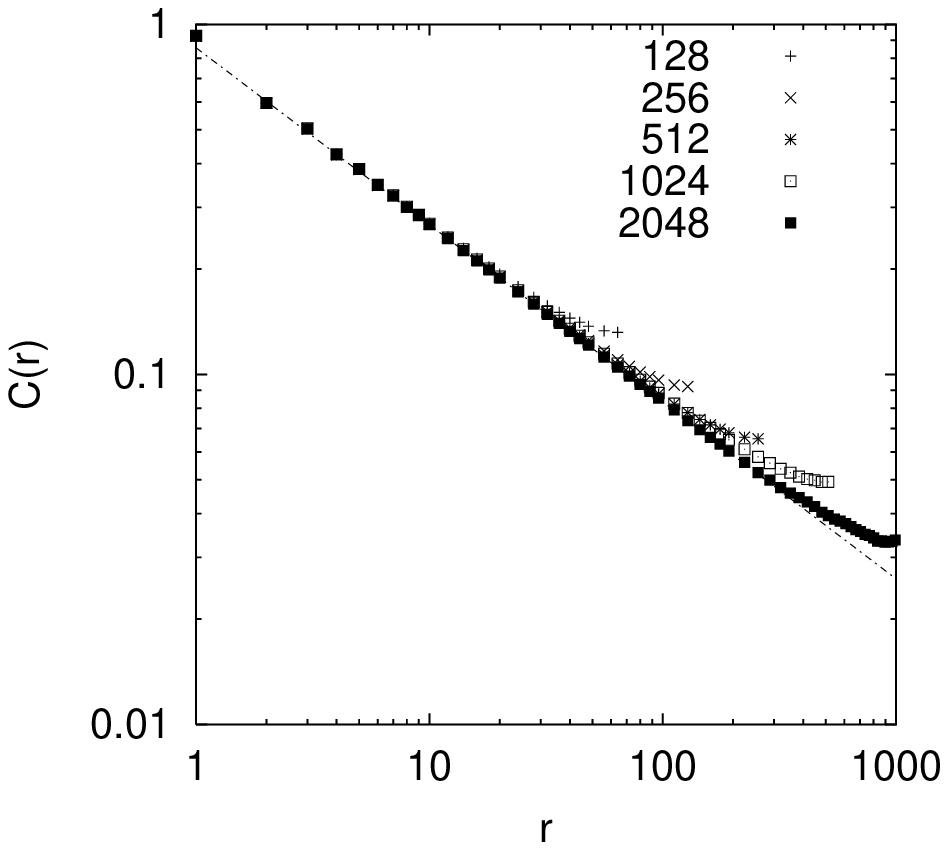}}
\caption{Log-log plot of $C(r),$ the magnitude of the spin autocorrelation
function, as a function of $r$ for $s=3/2.$ System sizes ranging from 
$L=128$ to $L=2048$ are plotted. The error bars are comparable to the 
point size. The data is consistent with a power law form, with exponent 
$-0.51.$ 
}
\label{fig2}
\end{figure}

Figure~\ref{fig1} shows the spin autocorrelation function $C(r) =
(-1)^r\langle s_j s_{j+r}\rangle$ as a function of $r$ for different
$L,$ for $s=1.$ The antiferromagnetic interaction down the chains causes
the oscillatory $(-1)^r$ factor. It is clear from the figure that the
correlation function does not decay exponentially.  As seen in the figure,
the autocorrelation function seems to decay as $1/r^\mu$ with $\mu \approx
0.68.$ However, motivated by the analytical considerations discussed in
Section 1, we try the functional form
\begin{equation}
C(r) = [A + B \ln r]^{1/2}{1\over{r^\mu}}
\label{log_spin1}
\end{equation}
with $\mu = 0.75.$ This is because the power-law part of $C(r)$ should
scale as $\sim 1/r^\mu$ with $\mu = 3/(2s + 2) = 0.75$ for $s=1,$
with possible logarithmic corrections from the marginally irrelevant
operator.  Accordingly, Figure~\ref{fig4} plots $C^2(r) r^{1.5},$
which should be a linear function of $\ln r.$ This expectation
is borne out by the plot. Logarithmic dependences are known to be
hard to distinguish from weak power laws, and the plots cannot be
used to choose between Eq.(\ref{log_spin1}) and the pure power law
decay of Figure~\ref{fig1}. (It would be even harder for the data to
discriminate between more subtle differences, e.g. Eq.(\ref{log_spin1})
with different exponents to the logarithmic term in the numerator.)
However, from Figure~\ref{fig4}, the data is certainly in agreement with
the analytical expectation.

\begin{figure}
\centerline{\epsfxsize=4in\epsfbox{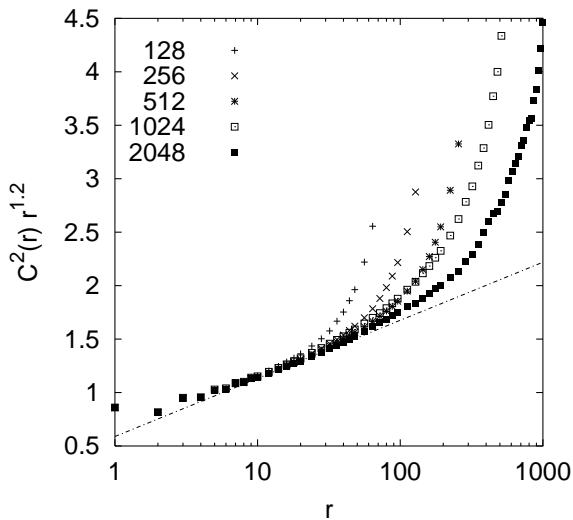}}
\caption{Plot of $C^2(r) r^{1.2}$ as a function of $\ln r$ for $s=3/2.$
System sizes from $L=128$ to $L=2048$ are plotted. The error bars are 
comparable to the point size.
}
\label{fig5}
\end{figure}

\begin{figure}
\centerline{\epsfxsize=4in\epsfbox{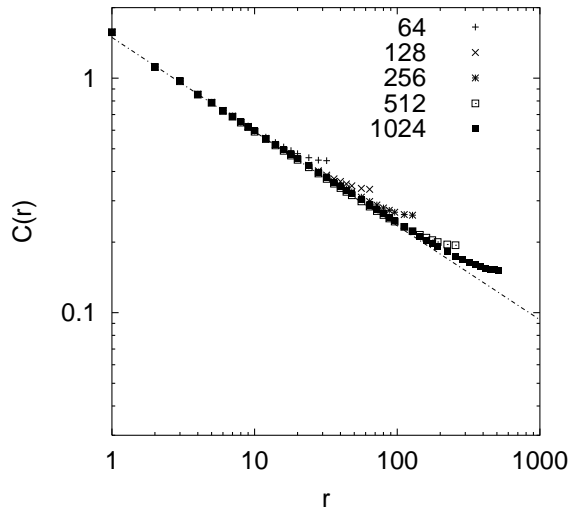}}
\caption{Log-log plot of $C(r),$ the magnitude of the spin autocorrelation
function, as a function of $r$ for $s=2.$ System sizes ranging from 
$L=64$ to $L=1024$ are plotted. The error bars are comparable to the 
point size. The data is consistent with a power law form, with 
exponent $-0.40.$ 
}
\label{fig3}
\end{figure}

\begin{figure}
\centerline{\epsfxsize=4in\epsfbox{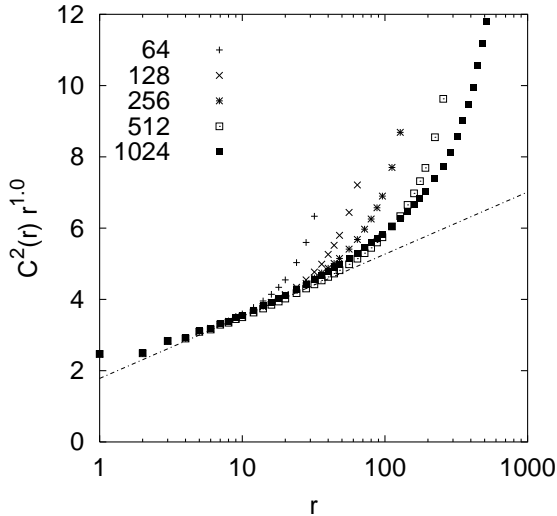}}
\caption{Plot of $C^2(r) r^{1.0}$ as a function of $\ln r$ for $s=2.$
System sizes from $L=64$ to $L=1024$ are plotted. The error bars are 
comparable to the point size.
}
\label{fig6}
\end{figure}

Figure~\ref{fig2} and Figure~\ref{fig5} are the counterparts of
Figure~\ref{fig1} and Figure~\ref{fig4} respectively for $s=3/2.$ Based
on the analytical prediction, Figure~\ref{fig5} plots $C^2(r) r^{1.2}$
as a function of $\ln r,$ since the power law decay part of $C(r)$ should
have an exponent of $\mu = 3/(2 s + 2) = 0.6.$ Likewise, Figure~\ref{fig3}
and Figure~\ref{fig6} are for $s=2,$ with $C^2(r) r^{1.0}$ plotted in
Figure~\ref{fig6}. Although the results for $s=3/2$ and $s=2$ are not
as clear as those for $s=1,$ we see that $C(r)$ definitely does not
decay exponentially for $s=2,$ or as $\sim 1/r$ for $s=3/2,$ the generic
behavior expected for integer and half-integer spin chains respectively.
The decay of $C(r)$ is consistent with the expectation that each $s$
corresponds to a WZW model of order $k=2s.$ The effective $\mu$'s from
Figures~\ref{fig2} and ~\ref{fig3} are 0.51 and 0.40 respectively.

\begin{figure}
\centerline{\epsfxsize=4in\epsfbox{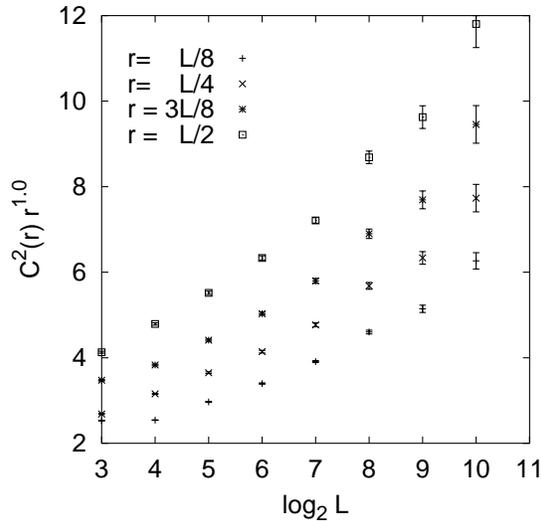}}
\caption{Plot of $C^2(r) r$ for fixed $r/L$ as a function of $\log_2 L,$
for $s=2.$ Four different values of $r/L$ are shown. The vertical bars 
are the error bars. There is a noticeable odd-even effect for $r=L/8$ 
and $r=3L/8$ for $L=8.$ For all the values of $r/L,$ the plots curve 
slightly upward.
}
\label{fig7}
\end{figure}

Since there is no clear linear region in Figure~\ref{fig6}, we also plot
the data in a manner that eliminates finite size effects. For finite
system size $L,$ the leading effect on $C(r)$ is to change its form from
Eq.(\ref{log_spin1}) to
\begin{equation}
C(r) = [A + B \ln r]^{1/2}{1\over{r^\mu}} F(r/L)
\label{fss}
\end{equation}
where $F$ is an unknown function. Since $\mu = 0.5$ for $s=2,$ we plot
$C^2(r) r$ for fixed $r/L$ as a function of $\ln L.$ The result should
be a straight line, with slope $B F^2(r/L).$ The results are shown in
Figure~\ref{fig7}. For all the values of $r/L$ shown, there is an upward
curvature to the plots. This is presumably due to subleading corrections
to the scaling form, since the leading correction is large enough to
change the effective $\mu$ in Eq.(\ref{log_spin1}) from 0.5 to 0.4.

\section{Conclusion}
In this paper, we have proposed a generalization of the Gutzwiller
wavefunction for spin ${1\over 2}$ chains that yields simple wavefunctions
for spin chains with $s>{1\over 2}.$ Remarkably, the spin spin correlation
functions for these wavefunctions have the same power law decay as
Wess Zumino Witten models with $k=2s,$ in contrast to the generic
expectation of $k=1$ for all half-integer spin and an exponential
decay for integer spin. This result was obtained through numerical
simulations for $s=1,{3\over 2}$ and 2, after mapping the model to a
classical statistical mechanical model with long range interacting spin
chains.  

In summary, we have taken free fermionic wave functions and found a
way of projecting them in a fashion that yields the WZW theory exponents,
a possibility that has been presaged in Ref(\cite{gogolin}). It remains
to be seen if there is a systematic way of finding Hamiltonians for
which the wave functions presented here are ground states.

\end{document}